# The Futility of Exoplanet Biosignatures

*Astrobiologists need to stop and ask themselves, what are we looking for when we look for life?*

**Authors**
Harrison B. Smith[1,*,✤] and Cole Mathis[2,3,✤]

**Affiliations**
[1]Earth-Life Science Institute, Tokyo Institute of Technology, Ookayama, Meguro-ku, Tokyo, Japan
[2]Beyond Center for Fundamental Concepts in Science, Arizona State University, Tempe, AZ, USA
[3]Santa Fe Institute, Santa Fe, NM, USA
*Author of correspondence: hbs@elsi.jp
✤Equal Contribution

## Abstract

The ultimate goal of astrobiology is to determine the distribution and diversity of life in the universe. But as the word "biosignature" suggests, what will be detected is not life itself, but an observation implicating a particular process associated with living systems. Technical constraints and our limited access to other worlds suggest we are more likely to detect an out-of-equilibrium suite of gasses than a writhing octopus. Yet, anything short of a writhing octopus will raise skepticism among astrobiologists about what has been detected. Resolving that skepticism requires a theory to delineate processes due to life and those due solely to abiotic mechanisms. This poses an existential question for the endeavor of life detection: How do astrobiologists plan to detect life via features shared between non-living and living systems? We argue that you cannot without an underlying theory of life. We illustrate this by analyzing the hypothetical detection of an "Earth 2.0" exoplanet. In the absence of a theory of life, we argue the community should focus on identifying unambiguous features of life via four areas of active research: understanding the principles of life on Earth, building life in the lab, detecting life in the solar system and searching for technosignatures. Ultimately, we ask, what exactly do astrobiologists hope to learn by searching for life?

# Life is not defined–What are we looking for?

How do we know life when we see it? We suspect that life beyond Earth will display a diversity of processes beyond what we see in our biosphere, but we don't know the limits of that diversity. Similarly, we suspect that abiotic processes on other worlds will display a diversity of possibilities but we don't know the limits of that diversity. This is a problem because current approaches to detecting life on exoplanets are based on identifying life's byproducts in atmospheres, but these byproducts are not unique to life ($O_2$ for example can be produced abiotically and biotically)[1]. If you want to truly know whether an observed feature is evidence of life, you need to have a clear definition of what life is, or at the very least an agreed upon method of verifying life detection claims via follow up experiments. If you want to sidestep the question of defining life, then you must reframe your question in terms of clearly defined features of life, and not the overarching concept of life itself.

One way that astrobiologists investigate the detectability of life on other planets is through complex planetary scale models[2]. But these models don't rely on any underlying theory of life, and instead consider specific sources and sinks of chemical species, and rules of their interactions. Separately, astrobiologists label these sources, sinks, or transformations as being due to life or non-life, tautologically defined by the fluxes they influence. For example, defining life via biotic fluxes of methane from methanogenesis[3,4] and defining abiotic fluxes via rates of serpentinization and impacts. These models can be used to create hypotheses about how fluxes and interactions lead to atmospheric observables, but assigning the cause of these observables to life or non-life is a choice of how life or non-life is labeled and constrained within the model. This means that models of exoplanet biosignatures are down-stream of theoretical decisions about what does and does not constitute life.

The question for astrobiologists is not "which features are associated with a specific set of fluxes?" but rather, "which features are *really* associated with life and non-life?" In order to answer this question, there is no way around committing to a working hypothesis of what we mean when we say we've detected life. Because *life* lacks a precise definition, unlike the way that a proton or water molecule are precisely defined, our options are:
I.  Define life detection as the observation of some combination or subset of all features we associate with known life.
II. Define life detection as the absence of observation of some combination or subset of all features we associate with known non-life.

Both options have their drawbacks. They require either high specificity (e.g. life is something which produces X, Y, and Z, as long as it's not in the presence of P or Q), or risk including combinations of features which aren't mutually exclusive to either life or non-life (e.g. life makes oxygen, but so do abiotic processes). Do we define life detection by overspecification or underspecification? This might seem like a hard choice, but only by over specifying can we pin down true-positive observations of life. Without true-positives, we can't have a "validation-set" of observations around which to forge hypotheses about life and make scientific progress in astrobiology[5].

Here we discuss how this reasoning implies that finding Earth 2.0 (one of the "holy grails" of exoplanet astrobiology research)[6], in itself, would be uninformative in our quest for discovering alien life. We then discuss the primary implication of this: that exoplanet biosignatures, as presently formulated, are generally futile if our goal is to detect life on any given planet with any confidence. We discuss the need for certainty in life detection, and conclude by offering alternative ways forward for making progress in our search to understand and discover life.

## Earth 2.0 is not life detection

Suppose it's 2042 and the successor to JWST identifies the atmospheric composition of a Venus-sized planet around a Sun-like star. Shockingly (or not), it contains detectable levels of $H_2O$, $O_2$, and $CH_4$; compounds which, in combination, are currently known to only exist in the presence of life[7]. Have we found life?

The problem with claiming we've found life on another planet, in the case of an "Earth 2.0" detection, is that there's no way to be certain it's actually *life* and not just some novel combination of abiotic factors that were previously unknown. This isn't unique to an Earth 2.0 situation, but is a general problem for life detection outside the solar system. We will be uncertain about the relationship between our observations (exoplanet atmospheres) and the processes they implicate (biotic or abiotic processes). This problem will exist for life detection claims within the solar system, and in artificial systems on Earth, but in those cases the ability to reduce our uncertainty via experiments is immediately available. In order to be certain about life detection claims, the biotic hypothesis needs to be the most parsimonious explanation of the observation. An example of what we mean by this is given below for a hypothetical detection of $O_2$ on a terrestrial planet (independent of any other atmospheric species).

The fact that abiotic processes and biotic processes produce $O_2$ in some conditions means that even a clear signal of $O_2$ is not a signal of any particular process–there are multiple known ways to produce $O_2$ on planets, both abiotically and biotically[8]. Our uncertainty about planetary scale processes on other worlds, combined with our uncertainty with how those processes could interact with atmospheres and surface materials, means that abiotic production of $O_2$ might be the most parsimonious explanation of an $O_2$ signal. Here by parsimony we mean a preference for simple explanations which can explain the observation, à la Occam's Razor[9]. But astrobiologists don't have a way to quantitatively compare the explanation that production of $O_2$ is abiotic compared to biotic. This is because we don't have a theory of living systems[10], nor a complete theory of abiotic planetary processes. So the evidence we will have in favor of a biotic explanation of this atmosphere would be the specific co-evolutionary history of Earth and its biosphere, and incomplete models of planetary surfaces and atmospheres. One of several possible abiotic explanations could be the existence of the common mineral rutile (titanium oxide) which may be able to catalyze the photolysis of $H_2O$ producing significant $O_2$ at the surface[11]. Is this a better explanation for our planet's atmosphere than life? There are statistical methods[12] (e.g. MDL, AIC, BIC) to compare the accuracy of models and their parsimony, but they require explicit mathematical definition of the models and predictions, which have not been developed for exoplanet life detection claims, with few notable exceptions[3]. Finding situations where we expect that $O_2$ to persist and coexist with other gas species might provide some constraints, but will not completely rule out abiotic explanations–simply due to our incomplete knowledge of chemical and geological processes (**Figure 1**). Our knowledge of the relative flux of known reactions has to be weighed against our uncertainty in understanding the extent of possible abiotic processes.

This will be true of all proposed biosignatures which have also been observed in abiotic systems, such as: phase separated boundaries[13], isotopic fractionations[14], morphological patterns in rocks[15], autocatalysis[16], and low-complexity metabolic byproducts[17]. Conversely, features such as high complexity molecules[18]; heredity, replication, and diversification[19]; and technosignatures cannot exist in abiotic systems. Features which cannot exist in abiotic systems constitute a different class of biosignatures that are not subjected to false positives because they are only associated with life in all known situations. The uncertainty associated with those features is a technical issue about detecting signals from other worlds, not a conceptual issue of whether the signal is associated with the underlying processes of life. This is distinct from an ensemble of features which have never been jointly detected in an abiotic system, but have been observed separately in abiotic systems. The existence of unambiguous biosignature suggests there should be an underlying theory of living systems which can explain why certain features can only exist because of life[10,18]. In the case of exoplanets we cannot rely on biosignatures which are not uniquely associated with life because we don't have a method to ground truth our claims beyond measuring atmospheric composition. It should be noted that even during the initial exploration of Mars via satellite remote sensing, Carl Sagan and colleagues concluded that without ground truthing, "interpretation of the significance of kilometer resolution features can only be based on first principles or on terrestrial analogy"[20,21]. If such conclusions were reached for Mars, how could we hope to be more confident for exoplanets?

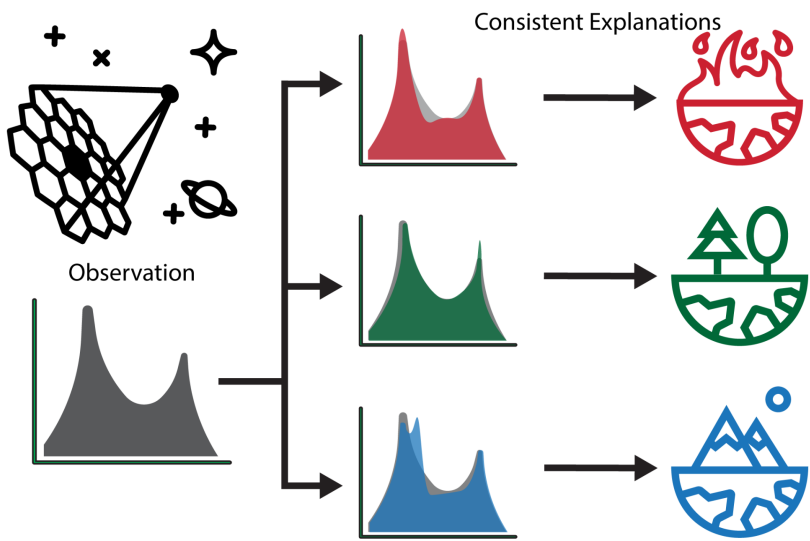

**Figure 1.** Different underlying planetary processes can lead to the same atmospheric observables, which means that a specific observation of an atmosphere cannot unambiguously identify the surface processes, including biological ones. Uncertainty in the types and scale of novel chemistry, planetary processes, and biotic processes amplify this problem.

Consider another example: There is an alien world that has been observing the $CO_2$ levels in Earth's atmosphere for a million years. What is their most parsimonious explanation for the meteoric rise in atmospheric $CO_2$ over the last 100 years on Earth, compared to the previous million years? Is it more parsimonious to attribute the cause to life, or non-life? If the aliens reason like us, they might be hesitant to attribute the changing $CO_2$ to life, even if this observable signature is inconsistent with their abiotic modeling. They might reason that the gap between their understanding and reality would be most parsimoniously explained by some misunderstanding of abiotic processes on planets. Perhaps they would assume the cause to be volcanic processes behaving in new and unexpected ways. Yet, that assumption would be wrong. The rise of $CO_2$ on our planet really *is* due to life–specifically, "intelligent" life excavating and burning extinct life[22].

There's a gut instinct to explain features on other worlds as being due to non-life where possible, because of the intuition that non-life is the "simpler" explanation. But it is only the simpler explanation if we assume that living processes are less likely than not–and this is wholly unconstrained. This may sound surprising at first, given that everything in the universe starts out as being abiotic (presumably). But we do not yet have a way to predict whether a system will tend towards living or non-living states[23,24]. This is, again, in part a consequence of not having a testable theory of life. Until we have a theory of living systems and non-living systems across environments, we are stuck with asking very specific questions about features of life, and features of non-life in very well understood environments.

Using Bayesian reasoning does not solve these dilemmas, because Bayesian inference (or any other kind of inference) is not a theory-free process (**Box 1**). In order to use data to make claims about the world (for example using a spectra to claim life has been detected) one first has to specify the relationship between the observation and the claim. Bayesian approaches provide a coherent mathematical framework within which we can reason about the strength of claims based on observations and prior assumptions, but doesn't say anything about the validity of the theories underlying these assumptions. For example, in astrobiology literature scientists commonly refer to a Bayesian equation assessing the probability of a biological explanation given some data[2]. But this probability is dependent on the prior probability of the biological model, and on the probability of that data within the biological model. Without a theory of living systems, the prior probability of biology in any particular environment is unconstrained, both numerically and conceptually. Without a theory of life you don't have a way to specify the phenomena you're assigning to life. This means that the probability of a biological explanation given the data is completely dependent on an unconstrained variable. In practice, this means that we cannot say which observations are more or less likely to be explained by life, since we have no way to constrain the likelihood of life.

**Box 1: Bayesian Life Detection is Not Infallible**

Bayesian inference is used to assess life detection claims by quantifying the probability of a biological explanation given some data (equation below). Consider two example life detection claims that are trivially equivalent in (un)certainty, despite one intuitively seeming more certain and constrained than the other.

- Jodie claims she detected life based on $O_2/CH_4$ observed in a terrestrial atmosphere.
- Matthew claims he detected life based on a random spectral signature from a random planet.
- Jodie says, "Hey wait, you can't just claim you found life–you need a model to back it up."
- Matthew responds, "I can just invent some new alien metabolism which produces what I want, and invent some new alien geological cycling which allows my metabolism."
- Jodie says, "But that metabolism and cycling isn't based on any known metabolism or cycling, so your model is inherently going to have lower certainty of claiming you found life. That is to say, your **P(L|D)** will be smaller than mine!"
- Matthew retaliates, "But where exactly in the Bayesian framework would that lower certainty be accounted for? I claim that it's lumped into the already unbounded **P(L)**. So my explanation doesn't *actually* have more uncertainty."

To further explain, let's keep in mind our Bayesian equation,

$$P(L|D) = \frac{P(D|L)P(L)}{P(D|L)P(L) + P(D|A)(1-P(L))}$$

where,
- **P(L|D)** is the probability of the biological model given the data
- **P(L)** is the prior probability of the biological model
- **P(D|L)** is the probability of the data given the biological model
- **P(D|A)** is the probability of the data given the abiotic model

Jodie's argument might be "my scenario A has more certainty around **P(D|L$_J$)**", where the subscript **J** indicates we are referring to Jodie's models. But really isn't it possible Matthew's scenario (subscripted **M**) could have just as much certainty around **P(D|L$_M$)**? Because Matthew can choose to construct those models in such a way that they will give an identical result of **P(L|D)** as a function of **P(L)**. To emphasize, in any scenario, **P(D|L)** is completely determined by whatever model you choose to use to represent your abiotic scenario and your biotic scenario, with no way to account for how "realistic" those are. Therefore, the only factor really determining **P(L|D)** would be **P(L)**, which is unconstrained (and theory dependent).

*Note:* In both scenarios we write the probability of the biological model given the data as **P(L)**, but really these are two different **P(L)**'s, since they are two different underlying models. In reality we intuitively recognize Jodie's **P(L)** as being "more likely" than Matthew's **P(L)**, but there is no way to confidently assign any different certainty to either.

## Without certainty, we can't learn new things about life, and so we can't enable unforeseen science

Even if we assume we can claim life has been detected in the Earth 2.0 example, we cannot use the information to improve any theories of life, because we haven't discovered anything that we did not know to be possible. In fact, we found exactly something which we've known for many years to be possible–because our planet has the exact same conditions. But, one might say, on our planet we know these conditions arise only because of life, so doesn't this give even stronger credence to the idea that Earth 2.0 harbors life? We could play along and say it does, but in that case it would only be a matter of linguistic convention–we don't know what caused the atmosphere we detected, but whatever it was we're going to call it life (this is analogous to the situation described earlier where modelers choose the labels they ascribe to different fluxes). Then the question would become: What would this teach us about life? How would it provide any information about what life is, what life does, or how we could learn more about life? It only tells us about the distribution of Earth 2.0 atmospheres. As to the nature of those atmospheres, it offers no answers. As to how one might follow up with more missions to understand the potential for life, it doesn't provide any kind of course correction or plan changes. The only course is to keep building bigger telescopes which can resolve more atmospheric data, more spatial resolution, and keep up distant-future planning to visit other star systems. Such observatories might one day be needed to test a yet-to-be-developed theory of life, but developing more capable observatories is not currently the limiting factor for our understanding of life in the universe–we are limited by the theories developed to assign cause to the observations we are making. How useful is the detection of a "very high confidence" biosignature if the biosignature doesn't teach us anything about life, or even change our approach to learning about life?

## Outside of enabling scientific progress, is there a reason we want to be certain about detecting life?

*"How sure we need to be before we accept a hypothesis will depend on how serious a mistake would be." –Rudner, 1953[25]*

It is up to the community to determine the implications of incorrectly making the claim, "life has been detected on another planet[26]." There are arguments to be made for both 'inconsequential' and 'catastrophic'. The 'inconsequential' side appears to have stronger evidence in its favor, as we have many examples of claimed life detection–in the mainstream media, by reputable scientists, and by reputable scientific organizations. For example, life has been reported as being discovered on the front page of the New York Times on at least three separate occasions, for at least two planets[27–29]. Even in the case of weaker claims of potential life detection[30,31], the claims have often acted as lightning rods for mustering up the scientific community to perform more rigorous analyses than might have been done otherwise, in the case where attention was not drawn to the claims. There are of course many other instances of life detection claims made all the time by less reputable sources.

However, scientists may face a much greater negative impact from simply asking questions which can lead to definite negative results than by falsely claiming positive results. The decrease in exobiology funding stemming from the negative results of the Viking life detection experiments led to the growth of an astrobiology community which has benefitted from "resisting closure to the question" of life detection, leading to a preference for asking ambiguous questions which cannot directly yield negative (or positive) results. Such a strategy may eventually backfire when clear results are never produced[20].

From the public's viewpoint, we know that a large fraction (sometimes even the majority) of the public polled already believes that there is life outside of Earth[32,33]. It's unclear if this has had any discernible impact on the public's view of science or science funding. Because this belief isn't linked to any scientifically agreed upon life detection claims (since the community currently agrees that life hasn't been detected on another planet), it appears that it may not be such an important claim after all to say that we've detected life on another planet. It may only matter if the detection leads to new information which significantly impacts the public or scientific spheres.

## How does astronomy work?

Given the arguments above, someone could reasonably ask how any scientific consensus on any field can be established[34], particularly fields in which direct experimentation are not viable options, such as astronomy, ecology and economics. Here we analyze the example of astronomy to help illustrate the way forward for exoplanet biosignature science. How is it that astronomers are able to make confident claims about the orbits and masses of planets around stars, or about the trajectories of meteorites and comets through space? In those situations direct verification of claims is impossible. Astronomers have no way to verify the location of planets centuries, or millenia in the past, or in the future. So how can we assign such a high confidence to their predictions? The reason is that the motion of planetary bodies is understood via the theory of gravity, which makes definite scalable predictions for a limited class of processes. It is applicable to experiments on Earth, extendable to predictions in the solar system, and defines what measurements impact predictions and which ones do not. Our theories of gravity are different from modern theories of biology because of their generality, which means we can easily deploy them to understand phenomena which occur very far away and in very different conditions. A theory of living systems suitable to make predictions about life on other worlds must also have this property—we must be able to use it to make predictions about life on Earth or in the solar system, and then understand how its predictions must change as we consider the diverse environments on other worlds. A theory of living systems must likewise clearly define which properties influence whether a system is living, and which properties are irrelevant for such predictions. The mathematical definition of gravity clearly does not depend on the color of objects, and clearly does depend on their mass. Likewise, a theory of living systems must make clear property dependencies. Hypotheses should be simple, specific, stated in advance, and have clear predictor and outcome variables[5]. This means that if *contextual* information is needed to understand biosignatures, it must be clear what exactly is needed.

# Four ways to learn about life

In the absence of a theory of life, one way to avoid these issues would be to build a telescope large enough to effectively "ground truth" life on exoplanets via very detailed surface features (such as forests, herds of elephants, or cities). Otherwise there are only four ways to learn more about life as a *phenomenon*, and make confident life detection claims (**Figure 2**):
  1. Studying life on Earth
  2. Finding reachable life in the solar system
  3. Making life in the lab
  4. Observing intelligent life via technosignatures

## Studying life on Earth

The most direct way to learn more about living phenomena is through our study of life on Earth. This can include expanding our notion of what the descendants of LUCA can and cannot do by studying specific and extreme environments[35], or it might include characterizing the large scale structure of biology to identify previously unknown patterns[36,37] which may recur in other biospheres[38]. This can involve anything from studying protein structures, to phylogeny, to metabolism, to population genetics, or studying ecosystem dynamics, or interactions between life and geochemistry. Systematic study of life on Earth may be sufficient to develop general theories of biological systems, or at least may help us identify ways to search for life on other worlds based on first principles[18,39]. Regardless, studying life on Earth expands the scope of features we know life *can* have, teaching us about the diversity of living processes.

## Finding reachable life in the solar system

The problems with detecting life outside the solar system stem from a variety of factors, but chief among them is that we won't be able to validate the origins of atmospheric features by direct verification of surface processes. This problem doesn't exist for life detection claims in the solar system, because missions to further test, or explore life hypotheses will be accessible on year to decade timescales. This means that an iterative process of exploration based on hypothesis generation will be possible. In that situation a life detection claim could be validated to the point where it is the consensus explanation among scientists because a diverse set of data will be available to support the explanation, and follow up experiments will be tailorable from the specific results from earlier missions. The technological, political, social, and economic coordination required to perform a similarly robust iterative process outside the solar system is such that we consider it impossible on the time scale of this century. Moreover, the astronomy community is already trying to build the biggest, most ambitious exoplanet observatories within the above constraints. Observations from these observatories will have little effect on the scale, function, or data collection desired from future generations of observatories (because the goal is always more ambitious, bigger, better telescopes).

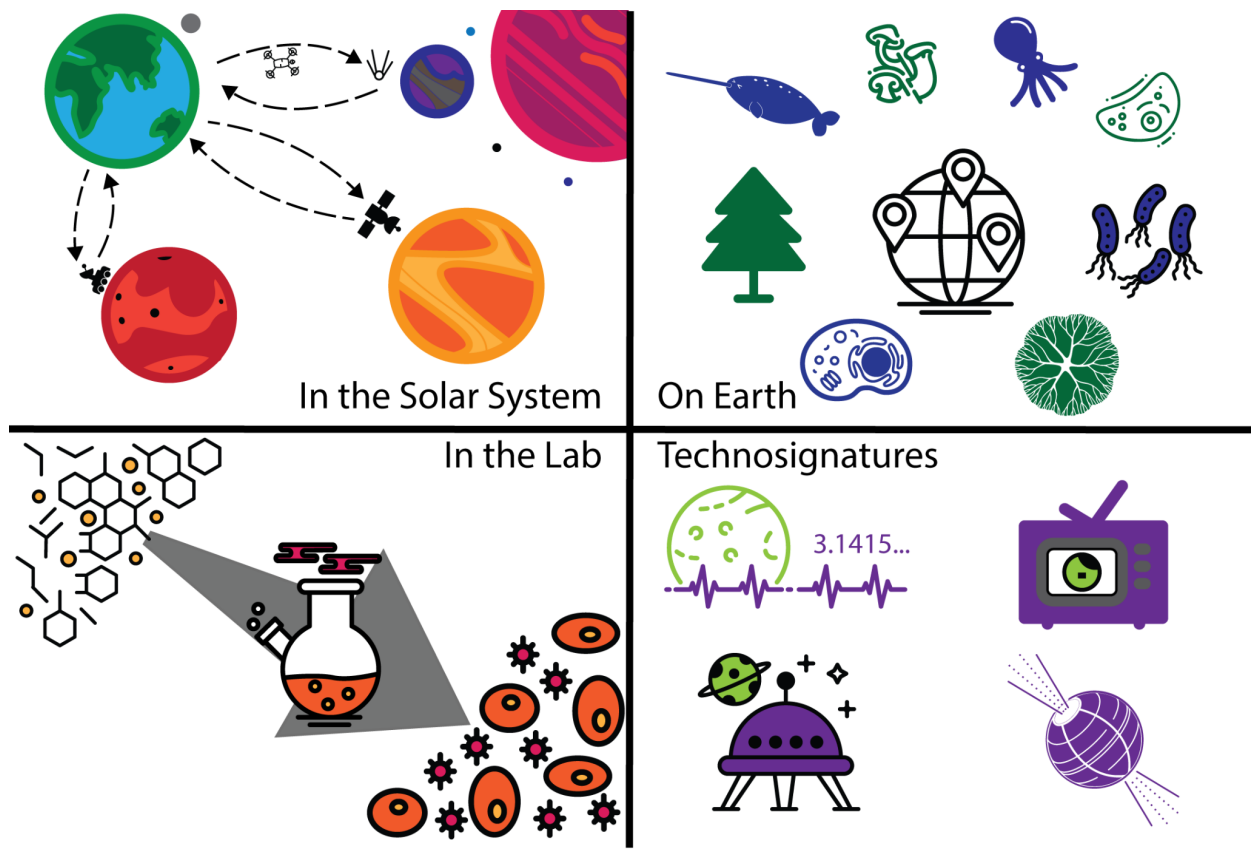

**Figure 2.** Different approaches to understanding the phenomenon of life in the absence of rigorous criteria for life detection on exoplanets. Life detection claims in the solar system can be driven by an iterative process that is tractable, placing them in a different class than exoplanet life detection. Studying life on Earth can help to identify the limits of known life and characterize general principles which may be applicable to all living systems. Building life in the lab enables direct experimental verification of the interplay between abiotic and biotic processes, which is not constrained by the historical contingencies of our biosphere. Technosignatures are representative of a class of unambiguous biosignatures which are uniquely associated with life.

## Making life in the lab

One of the reasons we are not able to formulate hypotheses about observable signatures of life on other worlds is that we have no empirical way to understand the limits of life's diversity. The life we observe on Earth is diverse in many ways but it is derived from a common ancestor. Astrobiologists anticipate alien worlds will display features which are unseen on Earth, but we are unable to understand the interplay between the richness of our imagination and the structure of physical reality. The most direct way to empirically explore the diversity of life beyond the descendants of LUCA is to develop experiments where we can observe the de novo emergence of life in the lab[10,40].

If we had access to experiments in which we could observe the dynamical consequences of animated matter, and where we could control the environmental conditions of those systems, we would be able to construct abiotic controls for those systems. The combination of abiotic controls and living systems would allow us to empirically investigate the consequences of living processes on the production of specific gasses (or ensembles of gasses) on the abiotic environment. The empirical results generated from those experiments would help develop a general theory of living systems, alleviating some of the issues described here, but more immediately it would provide empirical data about the relationship between living systems, their environmental context and observable consequences. These data could be used to generate hypotheses about what kinds of living processes are possible on different worlds and provide observational targets for them. In the absence of such data the only empirically motivated hypotheses about life on other worlds is constrained to the history of Earth life, which will (at best) lead to the detection of uninformative biospheres.

### Observing intelligent life via technosignatures

Why do technosignatures seem like foolproof evidence for life whereas $O_2$ in our Earth 2.0 example does not? It is because the very features that define a technosignature imply the existence of intelligent creators, and any alternative explanations that we could imagine *would be equally parsimonious* (alternatively, equally absurd). Say we received a transmission of an alien declaring the opening of the alien olympics. We would not be able to conceive of a more parsimonious explanation for this transmission than intelligent life. Radio waves have abiotic sources but Olympic broadcasts have no abiotic explanation. The evolution of aliens and the emergence of an alien olympics actually seem like the simplest explanation for our observed signature. This means that technosignatures represent an unambiguous biosignature that can be remotely detected. The challenges associated with their detection are in identifying and characterizing the signal (if it exists), not in determining whether or not it is due to life.

## Conclusion

This commentary is not meant to disparage the field of exoplanet biosignatures. On the contrary, it is intended to spur the community into thinking deeply about what exactly we learn from any given measurement, and especially the ensembles of measurements (e.g. $O_2/CH_4$) which the community widely views as indicative of life. When an international group of physicists set out to detect the Higgs boson, they were testing a specific hypothesis: If our current model of physics is correct, this particle collider should allow us to observe the Higgs boson. Following a positive detection, they started to measure many unknown properties of the Higgs boson that were uncharted territory for modern particle physics theory. What hypothesis (or hypotheses) is exoplanet biosignature science testing? And what new measurements would be enabled by finding a planet with an Earth-like atmosphere vs. finding a planet with an alternative atmosphere? What is gained by calling a detection of an Earth-atmosphere "life" if it doesn't provide any new information about the supposed underlying life, and would require follow up missions that are planned even in the absence of such a desired measurement?

As astrobiologists we believe the search for life beyond Earth is one of the most pressing scientific questions of our time. But if we as a community can't decide how to formalize our ideas into testable hypotheses to motivate specific measurement or observational goals, we are taking valuable observational time and resources away from other disciplines and communities that have clearly articulated goals and theories. It's one thing to grope around in the dark, or explore uncharted territory, but doing so at the cost of other scientific endeavors becomes increasingly difficult to justify. One of the most significant unifications of biological phenomena–Darwin's theory of natural selection–emerged only after Darwin went on exploratory missions around the world and documented observations. It's possible the data required to develop a theory of life that can make predictions about living worlds simply has not been documented sufficiently. But if that's the case we should stop aiming to detect something we cannot understand, and instead ask what kinds of exploration are needed to help us formalize such a theory.

## Acknowledgements

This work was supported by the Japan Society for the Promotion of Science (JSPS) KAKENHI Grant No. JP19K23459. CM would like to thank support from NASA through the Postdoctoral Fellowship Program. The views and conclusions contained in this document are those of the authors and should not be interpreted as representing the official policies, either expressed or implied, of NASA. The authors would like to thank Leroy Cronin and Michael L. Wong for feedback on the manuscript, and Sara I. Walker and Chris Kempes for insightful conversations.